\documentclass[sigconf]{nimeart}
\usepackage{xspace}
\usepackage{array}
\usepackage{booktabs}
\usepackage{tablefootnote}
\newcommand{\ra}[1]{\renewcommand{\arraystretch}{#1}}

\AtBeginDocument{%
  }

\setcopyright{cc}
\copyrightyear{2025}
\acmYear{2025}
\acmDOI{}

\acmConference[NIME '25]{International Conference on New Interfaces for Musical Expression}{June 24--27,
  2025}{Canberra, Australia}
\acmISBN{}
\newcommand{\MP}{\emph{Mimetic Poet}\xspace}
\newcommand{\MM}{\emph{Memetic Mixer}\xspace}

\begin{document}

\title{Mixer Metaphors: audio interfaces for non-musical applications}

\author{Tace McNamara}
\affiliation{%
  \institution{SensiLab, Monash University}
  \city{Caulfield East}
  \state{Victoria}
  \country{Australia}
}
\email{Tace.McNamara@monash.edu}
\author{Jon McCormack}
\orcid{0000-0001-6328-5064}
\affiliation{%
  \institution{SensiLab, Monash University}
  \city{Caulfield East}
  \state{Victoria}
  \country{Australia}
}
\email{Jon.McCormack@monash.edu}
\author{Maria Teresa Llano}
\affiliation{%
  \institution{University of Sussex}
  \city{Brighton}
  \country{UK}}
\email{Teresa.Llano@sussex.ac.uk}

\renewcommand{\shortauthors}{McNamara et al.}

\begin{abstract}
 The NIME conference traditionally focuses on interfaces for music and musical expression. In this paper we reverse this tradition to ask, can interfaces developed for music be successfully appropriated to non-musical applications?
  To help answer this question we designed and developed a new device, which uses interface metaphors borrowed from analogue synthesisers and audio mixing to physically control the intangible aspects of a Large Language Model. We compared two versions of the device, with and without the audio-inspired augmentations, with a group of artists who used each version over a one week period. Our results show that the use of audio-like controls afforded more immediate, direct and embodied control over the LLM, allowing users to creatively experiment and play with the device over it's non-mixer counterpart. Our project demonstrates how cross-sensory metaphors can support creative thinking and embodied practice when designing new technological interfaces. 
\end{abstract}

\begin{teaserfigure}
  \includegraphics[width=\textwidth]{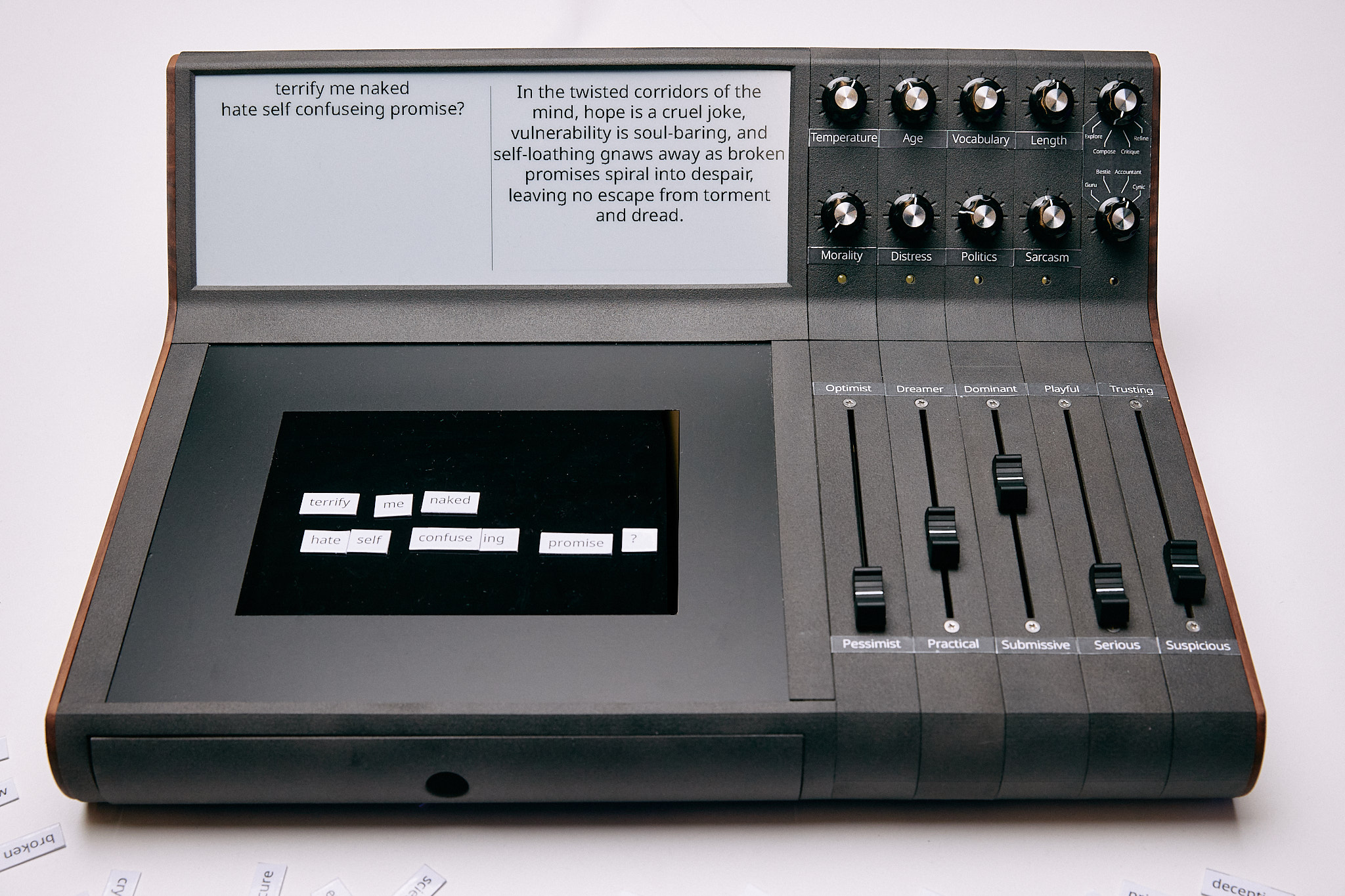}
  \caption{The \MM}
  \Description{The \MM - a tangible device for interaction with a generative AI system}
  \label{fig:teaser}
\end{teaserfigure}

\maketitle

\section{Introduction}
While there is a rich history of innovation and experimentation in the NIME community in developing interfaces \textit{for} musical expression and control, less research has been devoted to the application of musical interfaces to non-musical systems. Musical interfaces and their associated musical expressions are tangible, embodied, and interactive; strengths which are increasingly important when designing technology in any domain.\footnote{We acknowledge of course the broader field of tangible interaction design has addressed this area for many years \cite{shneiderman_direct_1983,ishii_tangible_1997,ullmer_emerging_2000,ishii_tangible_2008}.} To practically demonstrate how musical interfaces can enhance non-musical design, we draw on interface elements and metaphors from audio mixing and synthesisers to create the \MM---a bespoke tangible device for creative interaction with large language models (LLMs). Our work demonstrates how audio and music metaphors can serve as effective design tools beyond musical contexts.

Our application of musical metaphors comes from two perspectives. First, we consider how key paradigms from audio mixers and hardware synthesisers can be applied as conceptual tools by designers.  To do so, we outline the physical and functional affordances of faders, knobs, presets, filters and effects \cite{gibson_ecological_2014}. We then pose these audio norms as an exploratory framework for the design of non-musical interfaces that encourage expressive interactions, similar to what they do for musicians and audio engineers.  

Following this, we consider how musical metaphors may be beneficial to the users of non musical-interfaces. To evaluate how the use of music metaphors changes both perception and creative thinking, we built two versions of our tangible LLM device -- one with and one without the audio-inspired controls -- and provided them to a group of six fine-art graduate students in their shared studio space for a one week period for each version of the device. We found that the use of controls inspired by audio mixing and synthesiser design positively affected their conceptual approach, use and creative thinking about the device and its possible purposes. The immediate, direct and embodied control afforded by the audio-inspired interface allowed user's to creatively experiment and play with the device over it's ``non-mixer'' counterpart.

\section{Related Work}
\label{sec:related_work}
\subsection{Metaphor in Musical Interface Design}
The value of metaphor, widely used in HCI \cite{blackwell_reification_2006}, has been extensively applied to mapping functions for new, expressive musical interfaces \cite{waiteChurchBellesInteractive2016,grahamGestureEmbodiedMetaphor2014,dahlSoundBouncePhysical2010}.  Metaphor can be a useful tool to understand functions of innovative technologies which extend a musician beyond the bounds of traditional musical norms. For example,  \citet{gaddMetaMuseMetaphorsExpressive2002} used a rainfall metaphor to assist users in understanding granular synthesis. Others have focused on embodied and gestural metaphors in musical interaction design \cite{grahamManagingMusicalComplexity2015} and live performance \cite{wesselIntimateMusicalControl2002}. \citet{bauA20MusicalMetaphors2008} utilised musical metaphors at two levels of abstraction in their design process, the \textit{``instrument''} level focused on hardware and software specification, while the ``\textit{composition and interpretation}'' level entailed unique contextual interactions with their audio in-out device. They concluded the use of metaphor allowed for ``open-endedness'' in the design process. 

These examples demonstrate the benefit of metaphor in the design of musical interfaces has been established, both in the process of design and in user interaction. However, whether musical metaphors could be effective tools outside the disciplines of music interface design remains largely under-explored. 

\subsection{Musical Control Paradigms}
The intuitive control offered by audio mixers in particular has been a feature of HCI research \cite{kjeldskovSpatialMixerCrossDevice2020}. The gestural control mechanisms of audio interfaces are designed to facilitate ``tinkerability'': experimenting with alternatives, rapid shifts in directions and resets \cite{resnickDesignPrinciplesTools2005}. 
Many expressive control mechanisms from music and audio have become paradigmatic, each representing a unique type of control and creative function. There are no better examples of this than knobs and faders, classic gestural controls now common to musical interfaces. Recognising the importance of these mechanism in music interfaces more than a decade ago, \citet{gelineckQuantitativeEvaluationDifferences2009} performed a quantitative evaluation of the differences between knobs and faders, with both receiving above averages scores as ``intuitive'' and ``inspiring''.  \citet{gelineckStageVsChannelstrip2015} presented a comparative study of two other musical paradigms,  ``\textit{the channel-strip}'' and ``\textit{the stage}'' finding both to be effective  for a simple spatial panning task . More recently, \citet{rossmyButtonsSlidersKeys2022} did a survey on musical grid interface standards, focusing on buttons, faders and keys. They concluded there are clear conventions for these control mechanism in UI design and proposed that their work contributed to standardising these practised but yet to be formalised definitions. 

Extending beyond the musical domain, faders have even been considered philosophically in speculative epistolary fiction, warning against the risks of ``widespread Slider use and abuse'' wherein faders are used to control extremes of personalities \cite{mandikSliders2023}.
Others, such as \citet{morrisonEntanglingEntanglementDiffractive2024}   recognise the power of modes of musical interaction without the trepidation and instead argue for the need to further embrace interdisciplinary design within HCI. Though the strength of musical control paradigms have been recognised within this field of origin, there is much potential to apply these norms as design tools in other disciplines.  

In the following section, we discuss musical paradigms from audio mixing consoles and how they could be applied to other disciplines as functional and metaphorical tools.

\section{Mixer Paradigm}
\label{sec:mixer}
\begin{figure*}
    \centering
    \begin{tabular}{c|c}
         \includegraphics[width=0.45\linewidth]{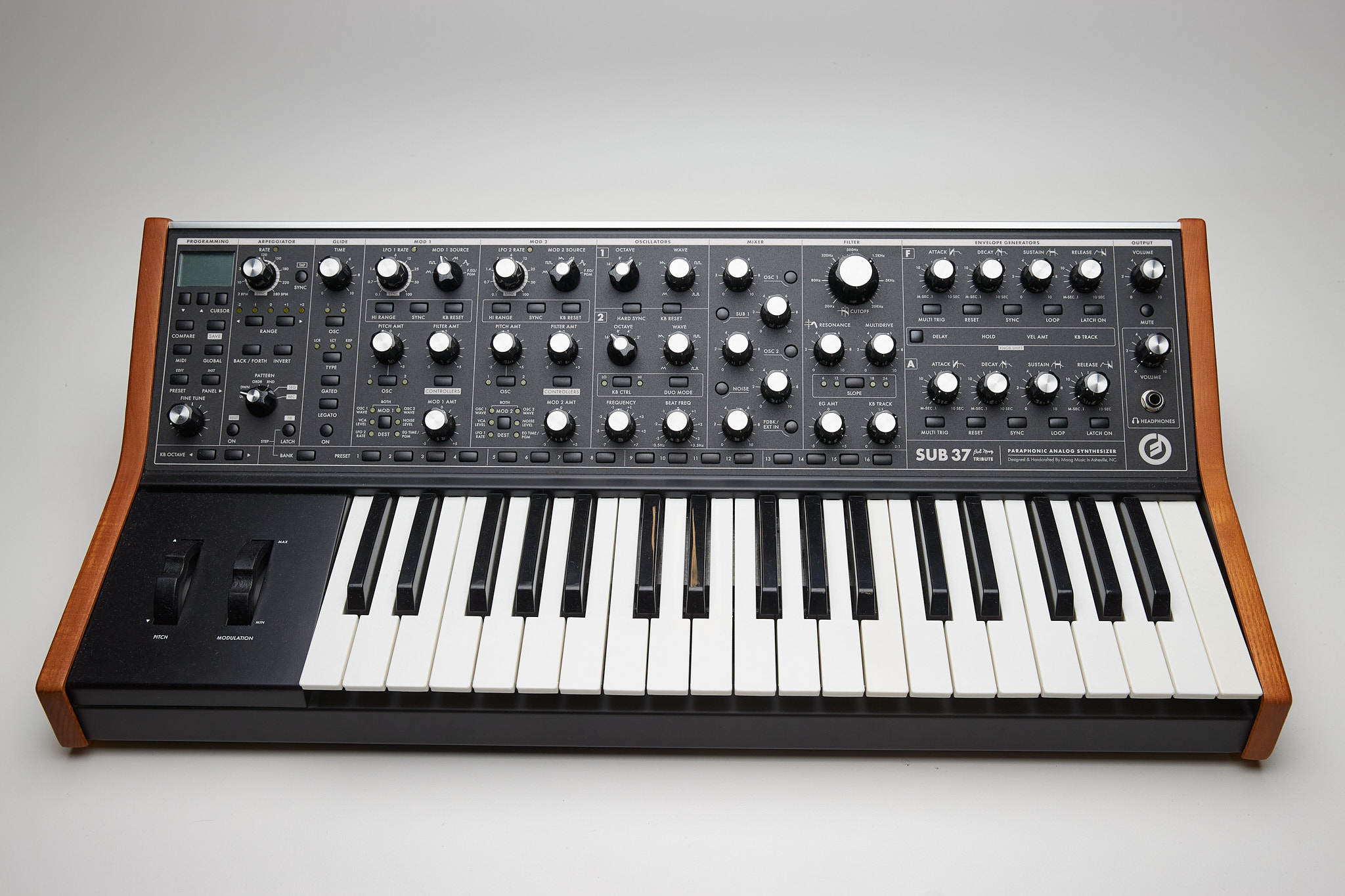} &
         \includegraphics[width=0.45\linewidth]{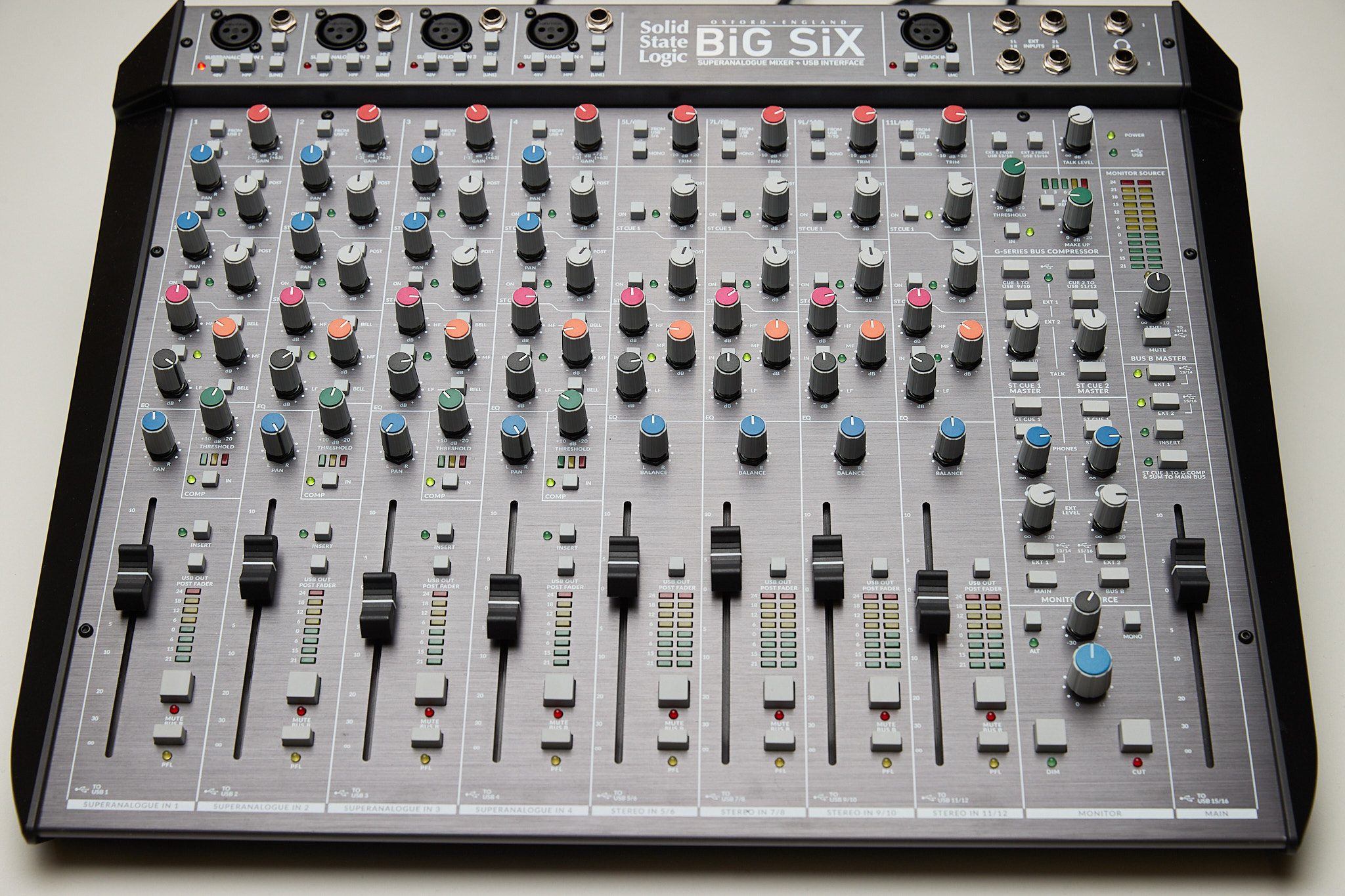}  
    \end{tabular}
    \caption{A Moog Subsequent 37 synthesiser with classic analogue aesthetic (left) and an audio mixer console with knob and fader controls (right).}
    % \description{An analogue music synthesiser with many knobs and controls (left) and an audio mixer with rows of knobs and large faders (right) }
    \label{fig:synths}
\end{figure*}

In this section, we describe control mechanisms from audio mixers that have become paradigmatic manipulation tools. For the benefit of designers from other fields who may not be familiar with audio mixer interfaces, we briefly describe their functional affordances and where relevant, sensory affordances of key control mechanisms.  We then provide a list of questions which help apply audio mixer paradigms to non-musical domains. Our aim is to present the mixer paradigm as a tool for divergent thinking to inspire new forms of expression for designers in both musical and non-musical fields. 

\subsection{Faders}
A foundation of audio mixer consoles, faders are linear control mechanisms which were invented to control multiple channels simultaneously \cite{izhakiFaders2023}. Commonly, there is a single function assigned per fader, like the volume control of an instrument. A single fader function can also be applied to a group of instruments or effects so the exact degree of change can be applied simultaneously to each channel. Faders may be bipolar, with two polar values either side of a neutral centre point, or polar, with a single sided spectrum or range \cite{rossmyButtonsSlidersKeys2022}. This simplicity plays a large role in the  intuitive nature of faders, or ``sliders'' as they are sometimes called. The position of the faders offers a visual representation of the internal parameter mapping providing the user to understand the mix in a glance \cite{coulesMixingEvolutionKey2020}. Faders have become synonymous with smooth transitions\textemdash fading in and fading out, allowing the control of a parameter free from the jagged edges of discrete time functions.  

Tom Dowd, scientist, audio engineer and inventor of faders, aimed to create a mode of fluid gestural control that would allow him to play a mixer console like a piano \cite{daleydanEngineersWhoChanged2004,silverstein_pioneers_2017}.
Faders allow sound engineers to manipulate unique points of expressive control under each finger, whilst moving each hand independently.  BBC broadcasting desks were originally designed so that the engineer would pull the fader towards themselves if they wanted to hear more of that instrument \cite{mellorBroadcastPracticeWhy2006,taylor1960sPyeBroadcast2021}.  Today, the action of turning up a parameter is mapped to sliding a fader vertically higher on a screen, or outwards on a physical console.

The following questions may serve as points of inspiration when considering how faders could be used as conceptual tools in a design:

\renewcommand{\labelenumi}{F.\arabic{enumi}}
\begin{enumerate}
\label{list:faders}
    \item \label{F1} What could be faded in or out?
    \item \label{F2} What parameters are being mixed together to create a whole?
    \item \label{F3} Which parameters could be assigned an independent fader to allow greater expressive control?
    \item \label{F4} What parameters could I group together with a single fader, to fade in and out together?
    \item \label{F5} Am I allowing both hands independent movement and expressive control?
    \item \label{F6} Am I allowing the fingers to used be independently and expressively?
\end{enumerate}

\subsection{Knobs}
Knobs are a versatile control mechanism common to audio and many non-musical fields. Knobs, or ``pots'' are rotary control mechanisms which offer the benefit of varying parameters based on rotational movement. Many knobs are labelled with numbers to indicate exact points of adjudgements. The concept of a knob to dial up the volume is now common on many sound producing apparatus. Knobs offer the possibility to pan between two binaries with a zero point in the centre or to dial up a single continuous parameter.  The later norm, encourages users to explore extremes unlike the nuanced and subtle control encouraged by  faders. ``Turn it up to eleven'' has become common vernacular after Spinal Tap (1984) parodied the musicians tendency to turn a knob as high as it will go. 

Knobs are also very intuitive as gestural control mechanisms.  Knobs are tweaked or dialled and our embodied understanding of these gestures conjures a very specific type of parameter adjustment. Most knobs fit neatly between two fingers and twisting them delicately to adjust a parameter creates a sensation quite distinct to sliding a fader. Additionally, knobs with discrete steps often provide a satisfying haptic feedback as the click into place. 

Inspired by the sensory and functional affordances of knobs in musical domain, the following questions may serve to translate these affordances to other disciplines:
\renewcommand{\labelenumi}{K.\arabic{enumi}}
\begin{enumerate}
\label{list:knobs}
    \item \label{K1} Which parameters could be expressively controlled in discrete increments?
    \item \label{K2} Could I offer creative control by panning between two binaries?
    \item \label{K3} What is the centre point between two binary extremes?
    \item \label{K4} Which parameters should be tweaked?
    \item \label{K5} Which parameters could be dialled up?
    \item \label{K6} What could I turn up to eleven?
\end{enumerate}

\subsection{Presets}
Many synthesisers and effects interfaces come with built-in sound presets which provide an automatic mapping to an internal matrix of parameters that developers have chosen as functional and aesthetic examples. Despite the integration of presets, synthesisers are designed to allow musicians to shape their own unique sound by manipulating each parameter to their taste. However, the presets of some synthesisers are so commonly used by musicians that many popular synthesises are defined by the sound of their presets.  

Presets are commonly activated by turning a selector knob and are given descriptive names to define the parameter mapping they represent. Using presets, a user may ``plug-in and play'', instead of setting each parameter independently before beginning. The immediate, no-fuss creative interaction offered by presets reduces the complexity of musical interface so as not to interfere with the potential for user flow states \cite{croninGrooveLessonsDesktop2008}. In one study, presets allowed inexperienced users to achieve similar results to expert users \cite{levaillantAnalyticVsHolistic2020}.  

The following questions can be used to consider how presets may be applied:
\renewcommand{\labelenumi}{P.\arabic{enumi}}
\begin{enumerate}
\label{list:presets}
    \item \label{P1} What quick start ``presets'' have I allowed the user?
    \item \label{P2} Which parameter mapping/setting of characteristics could be considered the aesthetic/tone of this work?
    \item \label{P3} Which four presets would best demonstrate the range and contrast in this system/work?
    \item \label{P4} What example/preset would best display the capabilities of this system?
\end{enumerate}

\subsection{Filters}
Filter controls, common to synthesizers and audio mixing consoles, allow certain frequencies to be sculpted away or selectively emphasised. Filters are often used as expressive tools for transitions and dynamics. Gradually opening up high or low frequencies will ``filter'' an instrument into a mix. Likewise, a sound source can be slowly removed by closing a filter and the frequency range it controls. Filters are applied on top of a sound source or instrument meaning the original sound can also be returned to. This means there is no need for the careful treading which may be necessary if there were risk of changing a sound irreversibly. 

The following questions may help to apply the concept of filters to other disciplines:
\renewcommand{\labelenumi}{$\nu$.\arabic{enumi}}
\begin{enumerate}
\label{list:filters}
    \item \label{v1} What aspects would I like to filter into/out of this work?
    \item \label{v3} How could filtering in a certain feature be used to create a transition?
    \item \label{v4} What could be filtered in and out to add more dynamic range?
\end{enumerate}

\subsection{Effects}
Effects refer to a range of sound manipulation techniques that shape timbre.  In some cases, these effects are fundamental to the defining sound of an instrument and are thus captured as part of the source sound. This means they can not be removed or changed in post-production. Commonly, effects are also added on top of a sound source to add colour, or other creative qualities. Effects can evolve a sound from basic and uninteresting to unique and dynamic. An effect can be applied to a single instrument or collectively to a group of instruments serving to unite their sound, and ``glue'' them together a space. Some commonly used effects in music production and performance are reverb, delay, distortion and chorus. 

The following questions may prompt ideas on how effects concepts can be applied beyond musical domains:
\renewcommand{\labelenumi}{E.\arabic{enumi}}
\begin{enumerate}
\label{list:effects}
    \item \label{E1} What effect could I apply to this element/work to morph it into something completely different?
    \item \label{E2} What effect could I apply to different elements to unify them?
    \item \label{E3} What element (ie.~adjective, feature, condition) could I treat metaphorically as an effect?
    \item \label{E4} How could I metaphorically apply (example effect: reverb, delay, distortion) to this work?
\end{enumerate}

We will now describe how we applied these audio mixer control norms as functional and conceptual tools in a tangible LLM interaction device. 

\section{The Memetic Mixer}
\label{ss:audio_mixer_design}

\begin{figure}
    \centering
    \includegraphics[width=0.8\linewidth]{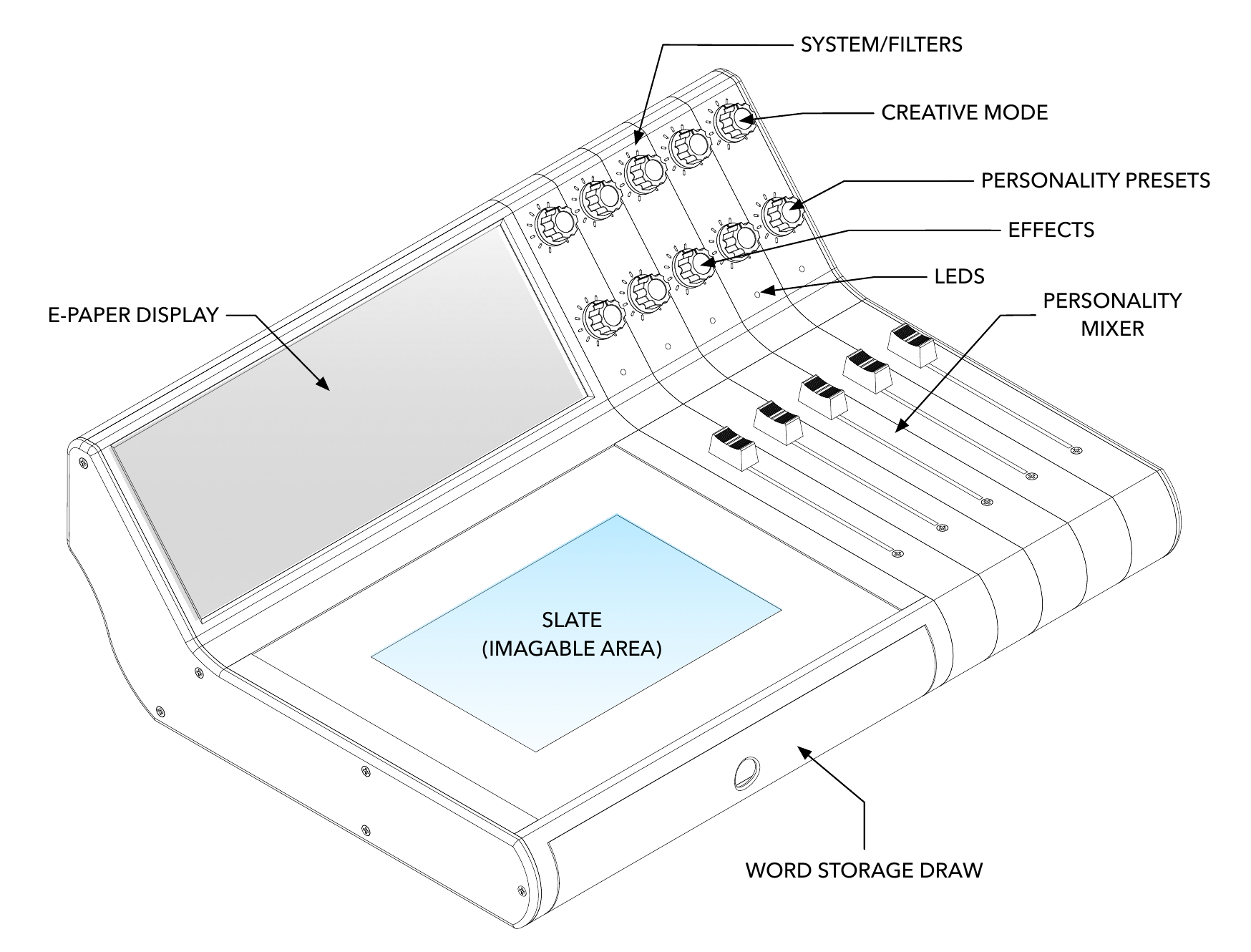}
    \caption{Schematic diagram of the \MM.}
    % \description{A wireframe drawing of the memetic mixer, showing a flat sensing area and an array of sliders and knobs on the right-hand side of the device.}
    \label{fig:mixer-xray}
\end{figure}

In this section, we offer an example of how we applied the audio mixer paradigm to LLM-interaction design, as a conceptual tool for design and to create greater expressive potential for users. Before introducing the details of the audio and synthesizer inspired controls, we briefly introduce the \MM device and application here. Further details, including technical implementation, are available in \cite{mccormack_mimetic_2024}.

Our motivation comes from exploring new ways of interacting with generative AI systems, in particular LLMs, where the dominant mode of interaction is the ``chat-bot'' style interface. While suitable for transactional modes of interaction, such as simple question and answer interactions, chat-bot interfaces emphasise the transitory, which quickly can become superficial. To encourage a slower and more considered mode of interaction -- something that felt more ``special'' and purposeful than a chat screen -- we settled on a physical design that used physical word ``tiles'' (inspired by \textit{magnetic poetry} and its well known ability to stimulate writer's creativity \cite{kapell_magnetic_1997}).

Words are placed on the device's slate (Fig.~\ref{fig:mixer-xray}) where they are recognised by the system and used to construct a complex series of internal prompt-chains \cite{wu_ai_2022}. The prompt chain is sent to the LLM, ChatGPT4-o in this case. The LLM response is then displayed on the device's e-Paper screen, along side the user input text as recognised by the device. It was necessary to design a means to manipulate the output-response without the cognitive demands and constraints typically imposed by prompt engineering. Hence the use of the ``mixer'' in the device.

Inspired by the modes of interaction used in analogue synthesisers and audio mixing technology (Fig.~\ref{fig:synths}), we conceptualised the \MM interface as a ``personality mixer''. In the same way an audio mixer is used to creatively mix together different audio tracks to create a final output, the \MM allows different traits to be fused into a personality for the LLM response (F.\ref{F2}).

Thus, audio mixing metaphors formed the conceptual model for a control interface to explore creative manipulation of the LLM's personality and behaviour. This design process began by drawing inspiration from analogue synthesisers like the Moog Subsequent 37 (Fig.~\ref{fig:synths}, left). Inspired by the analogue audio aesthetic, we added knobs and faders similar to an audio mixer (Fig.~\ref{fig:synths}, right).

We integrated five faders to create the LLM personality mixer (F.\ref{F1},F.\ref{F3}).  The personality traits were inspired by the ``Big 5''; five phenotypes recognised as the core characteristics of personality in psychology \cite{mccraeValidationFivefactorModel1987,digmanFiveRobustTrait1989,costaAgeDifferencesPersonality1976,goldbergStructurePhenotypicPersonality1993}. We refined these phenotypes to more intuitive labels based on more recent research on the Big 5, \cite{deyoungFacetsDomains102007,mccraeNEOPIMore2005,mccraerobertrClinicalAssessmentCan1986} and applying these personality categories to AI models \cite{cordobaInclusionLatentPersonality2012}. The personality mix is created by positioning the faders on axes of opposed personality traits which are as follows:\

\begin{figure}
    \centering
    \includegraphics[width=\linewidth]{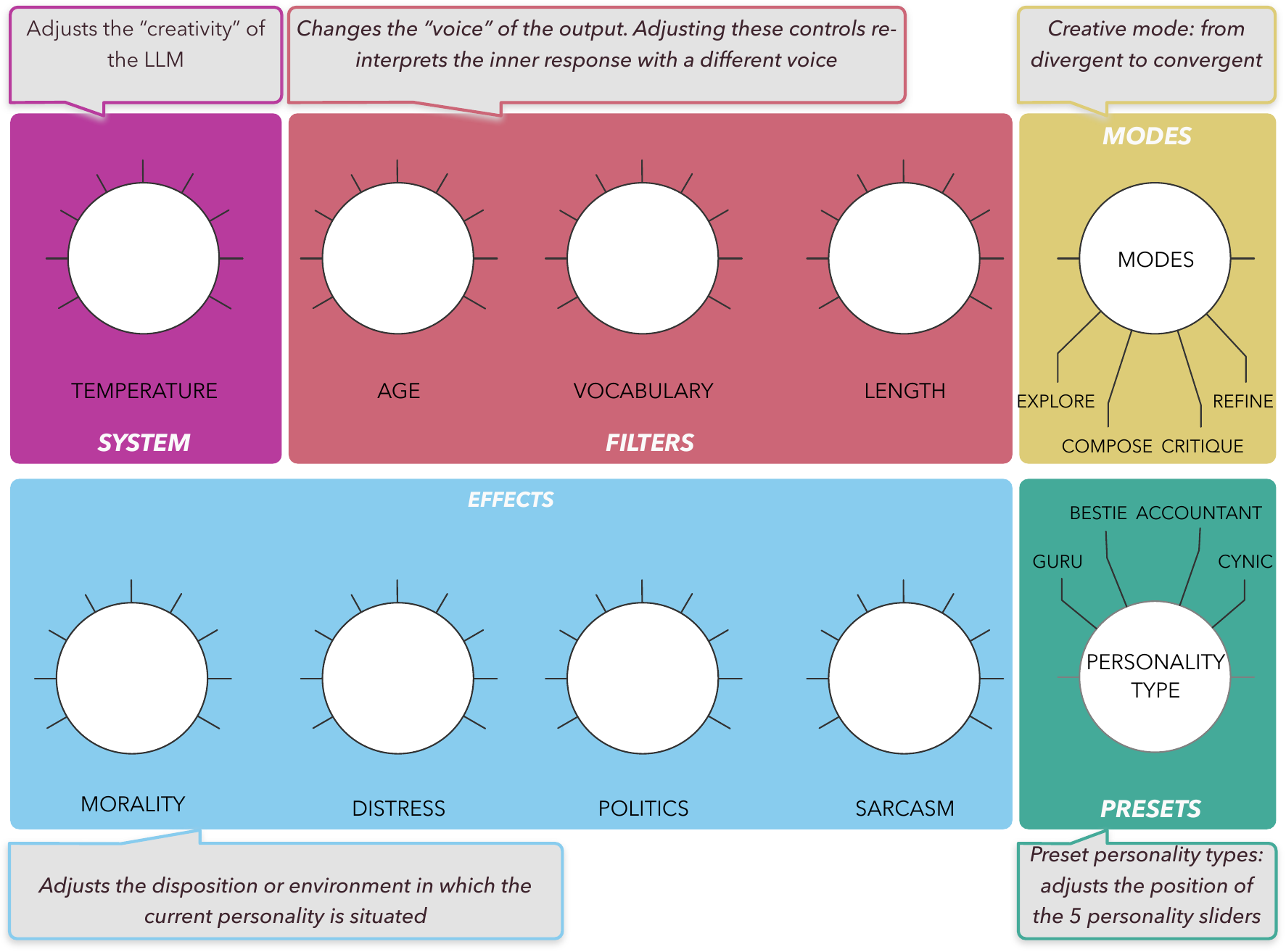}
    \caption{Knobs and their function groupings in the \MM}
    % \description{A graphic diagram showing two rows of four knobs and one selector on each row}
    \label{fig:knobs}
\end{figure}

\begin{quote}
\textbf{Optimist---Pessimist}\\
\textbf{Dreamer---Practical}\\
\textbf{Dominant---Submissive}\\
\textbf{Playful---Serious}\\
\textbf{Trusting---Suspicious}\\  
\end{quote}
Similar to the way an audio input is sent through distinct sections of a mixer console, each designated to particular functions, the knobs on the MM fall under the categories: \textit{system, filters, modes, effects, filters} and \textit{presets}  (Fig.~\ref{fig:knobs}).

The first preset knob allows the user to select one of four stages of creative process sending an internal prompt to instruct the LLM behaviour (P.\ref{P1}, P.\ref{P4}). The second preset knob was designed for \textit{personality presets} (P.\ref{P1}). The presets provide four contrasting personalities and tones of speech (P.\ref{P2}), which were chosen with the intention of creating personas that roughly covered the opposing spectres represented by the faders (P.\ref{P3}). Upon choosing one of four segments labelled with an archetypal persona, the motorised faders automatically move to the saved values representing that persona. A brief explanation of our chosen presets can be found in Table \ref{tab:presets}.

\begin{table}
\ra{1.3}
    \begin{tabular}{@{}p{0.15\linewidth}>{\raggedright}p{0.2\linewidth}p{0.2\linewidth}p{0.3\linewidth}@{}} \toprule 
         \textbf{Mode Presets}&  \textbf{Description}&  \textbf{Personality Presets}& \textbf{Description}\\ \midrule
         \textbf{Explore}& Divergent exploration of ideas & \textbf{Guru}& For toxic positivity look to the guru; the all-knowing teacher (and their ego)\\
         \textbf{Compose}& Convergent structuring of ideas & \textbf{Bestie}& The best friend who has complete faith in their counterpart but can always be trusted to be honest\\
         \textbf{Critique}& Feedback and evaluation&  \textbf{Accountant}& They may be down in the dumps but the practical accountant will always get straight to business\\
         \textbf{Refine}& Polishing and refining &  \textbf{Cynic}& Difficult to please and sure of themselves, the cynic will always play the contrarian\\
         \bottomrule
    \end{tabular}
    \caption{Memetic Mixer Preset Knobs}
    \label{tab:presets}
\end{table}

The filter knobs on the \MM were designed to add creative constraints to the LLM response without changing the personality mix or the mode preset ($\nu$.\ref{v1}). In this way, the user can explore variations of a single response as it changes according to the filter settings. We also integrated four effects knobs to ``colour'' the tone of the LLM output (E.\ref{E1}, E.\ref{E3}). Table \ref{tab:Knobs} outlines our choice of filter and effect labels. 

\begin{table}
\ra{1.3}
\centering
    \begin{tabular}
    {@{}p{0.2\linewidth}>{\raggedright}p{0.25\linewidth}p{0.15\linewidth}p{0.25\linewidth}} \toprule 
    \textbf{Filter}&  \textbf{Function}&  \textbf{Effect}&  \textbf{Function}\\ \midrule
    \textbf{Age}& baby to senior citizen&  
    \textbf{Morality}& malevolent to benevolent \\ 
    \textbf{Vocabulary}& very simple to extremely advanced &  
    \textbf{Politics}& liberal to conservative\\ 
    \textbf{Length}& one word to multiple paragraphs&  
    \textbf{Distress}& increases hysteria \\
    \textbf{Temperature}\tablefootnote{A native system parameter of many LLMs.}& increases uncertainty or randomness& 
    \textbf{Sarcasm}& increases sarcasm\\
    \bottomrule
    \end{tabular}
    \caption{Filter and Effects Knobs}
    \label{tab:Knobs}
\end{table}

\section{Comparative Study Design}
\label{sec:design_study}

\begin{figure*}
    \centering
    \includegraphics[width=\linewidth]{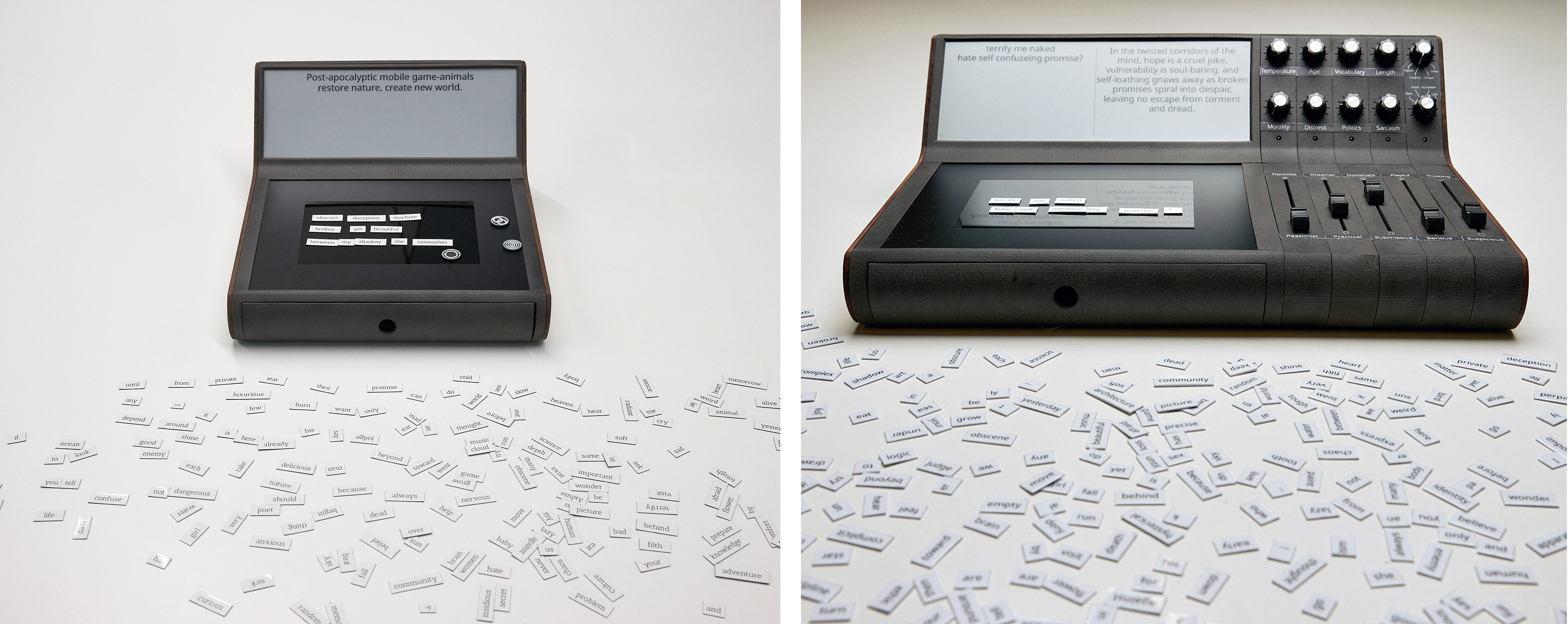}
    \caption{The \MP (left) a previous design without the use of audio-inspired controls, and the \MM (right), a re-imagined version of the \MP inspired by audio metaphors.}
    \label{fig:poet_mixer}
\end{figure*}

For the purpose of exploring if audio mixing metaphors can be used to facilitate more effective interaction with LLMs, we performed a comparative user study and analysis. This involved testing two tangible LLM devices; one which relied solely on word tiles and control markers to interact with an LLM (The \MP), and The \MM, described in the previous section, a new iteration augmented with physical controls knobs and faders to create a ``personality-mixer'' .

Both devices were tested in an ecological setting \cite{Brunswik1956,chamberlain_research_2012,benford_performance-led_2013} with six Fine Art Honours students (4 males, 2 females, age range 24-57). These students regularly worked from the same art studio where the devices were installed for the study upon consent. Being an ecological study, we did not add any participants from outside the art studio environment in order to inflate numbers. The number of participants was also suitable to allow each participant enough time to interact with a single device in a meaningful and sustained way.  

Participants attended an information session where they learnt about the functionality of the non-mixer device (The \MP) and how data would be used and collected.  Participants were not aware they would be evaluating another device in a following study and were initially asked to engage with the \MP alone.  While the study was carried out along side participants'  usual creative practice, they were free to use the device whenever and however they saw fit. In this way, our method drew from ecological studies \cite{Brunswik1956,chamberlain_research_2012,benford_performance-led_2013} which benefit from experiments taking place in a user's habitual environment rather than being obliged to test a device for a limited amount of time in a fabricated laboratory setting.

At the conclusion of the first study period using the non-mixer device, participants attended an in-person focus group consisting of semi-structured interviews and open discussion to gain insights from their experiences interacting with the device. The questions made no mention of the following study with the \MM. All comments were audio recorded and transcripts produced, along with researcher's notes and still photography. 

Following this, the same cohort of Fine Art honours students partook in another week-long study, approximately one month after the first. This time, interacting with the augmented personality mixer device (\MM) which was once again installed in the same communal area of their studio under the same conditions. Being familiar with the non-mixer device, participants were not given any new instruction on the use of the \MM, nor on the meaning or function of the additional controls, beyond the labels on device. 

At the conclusion of the second study period, we held a second 1-hour focus group, following the same format and protocols as the initial study. Each participant received two \$50 vouchers for participation, one following each focus group. Using a reflexive, inductive thematic analysis approach \cite{braun_using_2006,braun_thematic_2022}, the interview transcripts and notes were independently coded by two members of the research team to draw out the significant themes from both studies. We summarise the results of this analysis in the next section. The studies were approved by our university ethics committee.

\subsection{User experience with non-mixer device}
In discussion with participants about their experience throughout the week using the non-mixer device (\MP), their appreciation of the tangible nature of the device was apparent. One participant remarked, \textit{``the physicality of it is a nice touch, especially because it is an AI model''}. Another expressed approval of the \textit{``analogue style''} of the interface.  The importance of physicality was cemented by a claim that this was in fact the only appeal of the device: ``\textit{If it was completely digital, it wouldn't have [any] attraction to it.}'' However, the physicality of the device alone was not enough to sustain the participants' interest.  One reported, \textit{``I got bored of [it] pretty quickly''} which was supported by another participant who admitted, ``\textit{I couldn't do a sustained session with this.}'' 

The rapid decline in engagement seemed connected to a lack of expressive range in the interactions. One user imagined a more appealing device wherein, \textit{``maybe you had a variation of things and you weren't a hundred percent sure what they mean would probably keep you into it a little bit longer and maybe you'd get more of those iterations and stuff happening.''}

Participants wanted the LLM to expand upon their ideas with divergent associations, not just redress the same ideas with synonyms. One expressed, \textit{``I was being quite associative. And it would have been fun if it came back at me quite associative as well. With just, like, a few other words that were not the same.''}  Cross-disciplinary correspondences were mentioned as a way to incorporate divergent sources for ideas: \textit{``taking what you've written and transposing it into like a different discipline. And it might even pick a discipline\ldots randomly. This is what an architect would think.''} Another participant elaborated, \textit{``It's like how they say to us as creatives \ldots `go and consume other art', or if someone's a dancer and they do ballroom\ldots go and learn river dancing or something, because that's going to have an impact on your creative output. So it's like dropping it into something that's completely different\ldots And that's probably what. Maybe it was, like, missing a bit.''}

The desire to draw on ideas from other disciplines was reiterated throughout several points of the discussion. A creatively satisfying interaction with a LLM was envisioned as 
\textit{``almost like a brainstorming session''} wherein it could provide \textit{``unconventional combinations of \ldots things.''} The ability it iterate upon ideas was perceived as a straight of AI models. One participant elaborated, \textit{``I always think about AI as\ldots limitless in its ability to conceptualise stuff. So you kind of want to see what it's going to come up with.''} This was confirmed by one participant's want for an interaction \textit{``that might reveal something that you just would have never occurred to you because you're already in this fixed way of thinking. And that's what I would be using it for, to help spark some other thoughts that I could then go away with myself.''}

Interestingly, participants began to develop their own cross-disciplinary metaphors in imagining the type of device which could be more creatively satisfying. One user expressed their desire for \textit{``an ugly or extreme sound''}, drawing on an auditory metaphor to describe a more stimulating response. Another participant suggested the LLM could be manipulated with tactile \textit{``knobs and sliders''}. They imagined a knob could be used to \textit{``dial up or down''} personality traits such as \textit{``snarkiness''} or \textit{``verbosity''}. These comments indicate that modes of interaction with audio were culturally established in our participants, being a standard reference for intuitive control styles even amongst non-experts. 

\subsection{User experience with mixer device}
In recounting their experiences interacting with the \MM augmented with knobs, faders and presets, participants perceived a notable difference in the potential range of expression. There was a new found ability to experiment with different ``mixes'' of the LLM response, to \textit{``morph it through the effects pedals, see what else happens.''} The filter and effects knobs allowed participants to explore different tones of LLM response. One shared, \textit{``there was a nice kind of play where you could just put your input in and then just keep dialling\ldots like setting something and then just dialling one thing up or down. Yeah, definitely had a greater impact.''}

In exploring the expressive capabilities offered by the mixer controls, some participants preferred a ``kind of fine tuning.'' This process of \textit{``refining the refining''}, as described by one user, allowed users to make precise and nuanced changes to the LLM response. Another participant however found that \textit{``it was more interesting when the parameters were set to extremes,''} seeing the obvious changes in tone as \textit{``more fun and interesting''}. Experimenting with the results from different parameters also provided incentive for continued interaction. As one participant put it, \textit{``you're still wanting to work with quite strict parameters to then see the unpredictable predictability to be interesting''}.

The participants found the idea of a personality mix \textit{``intuitively makes sense for those sliders''.} They were not inhibited by the lack of instructions for these new control parameters as the audio interface metaphor made the interaction style immediately apparent. One participant expressed, \textit{``I kind of like that\ldots there's no instructions,''} while another affirmed, \textit{``to be clear, I would not read the manual!''}  Participants believed an instruction manual was not necessary for the exploratory and open-ended style of interaction facilitated by the audio control mechanisms. One elaborating on this idea stating, \textit{``having instructions or a manual, like would be very relevant if you're using this, like, very intentionally, like\ldots towards whatever project you're working on. I think just for this, it was good in a way not to''}.

In general, participants believed the audio mixer metaphor allowed for a more \textit{``malleable discussion''}, wherein the mixer offered more expressive control over the LLM output. There was a sense of satisfaction in using the controls to \textit{``shape it in front of you,''} which was not possible in the non-mixer iteration of the device. 

\section{Discussion}
User experiences with the non-mixer LLM device made evident that the word tiles alone did not provide incentive for ongoing interaction. This problem was largely related to the predicable nature of the LLM responses which did not expand on user input in the way of \textit{``brainstorming''} or providing \textit{``unconventional combinations''} of concepts. It was interesting to observe that participants suggested that cross-disciplinary connections (\textit{``transposing it into\ldots different discipline}'') would potentially offer more creative value to the interaction. At this stage of the study, the \MM did not exist.  This comment was completely unprompted and stemmed directly from the desires of the participants themselves evidencing that they too recognise the value in cross-disciplinary design. 

Just as cross-disciplinary associations arose as a suggestion to improve the non-mixer device, participants also spontaneously began to use sound and audio control metaphors to describe a more fruitful interaction.  One participant's desire to be able to ``\textit{dial up or down}'' personality traits (K.\ref{K5}) and another's suggestion to integrate ``\textit{knobs and sliders}'', suggests that the parametrisation of these gestural controls has become paradigmatic outside of their use in audio and music. 

Comparatively, users were decidedly more positive in describing their experience using the \MM device. Participants were enthusiastic about the increased level of control over the LLM output they were offered by the mixer control mechanisms. The distinct affordances of knobs and faders allowed for both an exploration of ``\textit{extremes'}' (K.\ref{K3}) and a more nuanced ``\textit{refining''} (K.\ref{K4}). This suggests that the audio mixer controls were effective in allowing a spectrum of control which was suited to a range interaction modes. 

There was the potential that users unfamiliar with audio and music paradigms may not have grasped the mixer metaphor, or found the control mechanisms unintuitive. However, this fear was calmed by the adamant statement that users did not feel the necessity for a device manual, nor would they read it if one was provided. They were of the opinion that a personality mixer ``\textit{intuitively makes sense''}.  

The effectiveness of audio paradigms as conceptual tools was also revealed through the participants' (unprompted) use of music metaphors to describe their experiences with the \MM despite the lack of any actual sound or auditory affordances in the design. Participants spoke of language as akin to ``chords'' which they could ``\textit{morph through\ldots effects pedals},'' rather than describing the output editing process using technical or dry language. These sonic descriptions of language suggest that mixer metaphor was effective enough for the participants to carry the analogy into their own language use.  

Comparing the interviews about user experiences with the \MP (non-mixer device) and the \MM made clear that the personality mixer metaphor greatly improved the participants' experience interacting with the LLM. The personality mixer metaphor was functional in allowing a greater control of the LLM response and also gave users an embodied agency in manipulating the text output with gestural knob and fader controls. 

This study did not focus on a goal-oriented interaction with the \MM but allowed users the freedom of open-ended exploration.  In future work, we believe it would be valuable to explore whether the personality mixer metaphor may be useful for users with more specific creative goals. It may also be interesting to explore whether the personality mixer is equally appealing to creatives who work in audio and musical fields, or whether the cross-disciplinary metaphors cause cognitive tension to arise. 

\section{Conclusion}
\label{sec:conclusion}
In this paper, we have demonstrated the application of audio and music metaphors as creative design tools for interaction with non-musical devices. Beginning from the perspective of the designer, we outlined the functional affordances of control mechanisms common to audio mixer consoles: faders, knobs, presets, filters and effects. We then offered a template of questions to assist in translating each control paradigm to non-musical disciplines as design metaphors. 

Following our exploration of musical metaphors in the design process, we assessed their effectiveness in user interactions. We undertook two, week-long ecological studies wherein users explored using a tangible LLM device, first without, and then with an audio-mixer control interface. Our study found the audio-mixer control mechanisms not only afforded participants more nuanced control of the LLM output, but also allowed them to consider language metaphorically as a sound they were shaping. This lead to divergent ideas and an increase in multi-sensory analogies which are associated with creative activity. The music and audio metaphors created a more satisfying style of interaction for users and increased incentive for ongoing interaction with the device.

Overall, applying metaphors and physical control mechanisms from audio and music to other disciplines proved effective in guiding our creative process as designers.  It also allowed users an expressive style of interaction with the LLM device due to the embodied nature of the gestural controls and the cross-sensory metaphors they inspired. 

It is evident that applying knobs, faders and effects will not make an interface musically expressive if it was not designed for this function. However, any interface with adjustable parameters can be considered \textit{through the lens of audio mixing} so as to foster new perspectives and draw on the creative strengths of the musical domain. We would encourage other designers to consider how paradigms from musical norms could be applied in their work. We believe musical expression has much potential to enrich other fields for its conceptual and functional fullness.

%% The Ethical Standards section is mandatory for all NIME submissions.
\section{Ethical Standards}

Our project received competitive funding support from our federal government's main research funding body. Conditions of funding require that all research undertaken with the funding meets our country's responsible standards of professional research and ethics, which we followed at all times in undertaking this research.

Our user study with human participants was approved by our university ethics committee and participation was opt-in and voluntary. Participant's were free to opt-out at any time and the true nature and purpose of the study was made clear to participants before seeking their consent, as was the data collection and privacy requirements. Participants received a small financial compensation for proving their time and opinions over two one-hour focus group sessions.

No individual personal data, apart from each participant's name, gender and age was collected from participants and data sent to OpenAI's servers used a role account, not linked to any individual participant or person. The method of communicating with the LLM -- using a small vocabulary of fixed word tiles -- meant that no personally identifiable information was sent to the LLM.

As acknowledged in Section \ref{sec:conclusion}, the visual nature of the device raises issues of accessibility, for example to the blind and low-vision (BLV) community. We outlined how we would seek to address this issue in further developments of the research.

%To ensure objectivity and transparency in research and to ensure that accepted principles of ethical and professional conduct have been followed, authors must include a section “Ethical Standards” before the References.
%This section should include (if relevant): information regarding sources of funding, potential conflicts of interest (financial or non-financial), informed consent if the research involved human participants, statement on welfare of animals if the research involved animals or any other information or context that helps ethically situate your research.
%For help with the ethics section, feel free to ask on the NIME forum: \url{https://forum.nime.org}.

\begin{acks}
This research was supported by an Australian Research Council grant DP220101223. The \MP and \MM hardware was built by Elliott Wilson.
\end{acks}

\bibliographystyle{ACM-Reference-Format}
\bibliography{NIME-refs}

%\appendix

\end{document}